\input{aipcheck}
\documentclass[
    ,final            
  ]
  {aipproc}

\layoutstyle{6x9}

\begin{document}

\title{Relativistic Mean-Field Models and Nuclear Matter Constraints}


\author{M.~Dutra}{
  address={Departamento de F\'isica, Instituto Tecnol\'ogico de
Aeron\'autica-CTA, 12228-900, S\~ao Jos\'e dos Campos, SP, Brazil}
}

\author{O.~Louren\c co}{
  address={Departamento de F\'isica, Instituto Tecnol\'ogico de
Aeron\'autica-CTA, 12228-900, S\~ao Jos\'e dos Campos, SP, Brazil}
}

\author{B.~V.~Carlson}{
  address={Departamento de F\'isica, Instituto Tecnol\'ogico de
Aeron\'autica-CTA, 12228-900, S\~ao Jos\'e dos Campos, SP, Brazil}
}

\author{A.~Delfino}{
  address={Instituto de F\'isica, Universidade Federal Fluminense, 
24210-150, Boa Viagem, Niter\'oi, RJ, Brazil}
}

\author{D.~P.~Menezes}{
  address={Departamento de F\'isica, CFM, Universidade Federal de Santa
Catarina, CP. 476, CEP 88.040-900, Florian\'opolis, SC, Brazil}
}

\author{S.~S.~Avancini}{
  address={Departamento de F\'isica, CFM, Universidade Federal de Santa
Catarina, CP. 476, CEP 88.040-900, Florian\'opolis, SC, Brazil}
}

\author{J.~R.~Stone}{
  address={Oxford Physics, University of Oxford, OX1 3PU Oxford, 
United Kingdom}
  ,altaddress={Department of Physics and Astronomy, University of Tennessee, 
Knoxville, Tennessee 37996, USA}
}

\author{C.~Provid\^encia}{
  address={Centro de F\'isica Computacional, Department of Physics, University
of Coimbra, P-3004-516 Coimbra, Portugal}
}

\author{S.~Typel}{
  address={GSI Helmholtzzentrum f\"ur Schwerionenforschung GmbH, Theorie,
Planckstrasse 1,D-64291 Darmstadt, Germany}
}

\begin{abstract}
This work presents a preliminary study of $147$ relativistic mean-field (RMF)
hadronic models used in the literature, regarding their behavior in the nuclear
matter regime. We analyze here different kinds of such models, namely: (i)
linear models, (ii) nonlinear $\sigma^3+\sigma^4$ models, (iii)
$\sigma^3+\sigma^4+\omega^4$ models, (iv) models containing mixing terms in the
fields $\sigma$ and $\omega$, (v) density dependent models, and (vi)
point-coupling ones. In the finite range models, the attractive (repulsive)
interaction is described in the Lagrangian density by the $\sigma$ ($\omega$) field. 
The isospin dependence of the interaction is modeled by the $\rho$ meson field.
We submit these sets of RMF models to eleven macroscopic
(experimental and empirical) constraints, used in a recent study in which
$240$ Skyrme parametrizations were analyzed. Such constraints cover a wide range
of properties related to symmetric nuclear matter (SNM), pure neutron matter
(PNM), and both SNM and PNM. 
\end{abstract}

\maketitle



Quantum Hadrodynamics (QHD) is an approach based on quantum field
theory much used in the description of hadronic environments, such as nuclear and
neutron matter. It is based on local Lagrangian densities whose free parameters
are adjusted in order to reproduce basic nuclear matter bulk properties at zero
temperature. In general, nuclear matter is well described by different versions
of the relativistic mean-field (RMF) models constructed via the QHD approach.

In order to select a set of RMF models which better describe nuclear matter
properties, as well as the physics of pure neutron matter, we submit $147$
relativistic parameterizations to 11 constraints, out of which 4 are related to
symmetric nuclear matter, 2 to pure neutron matter, and 5 related to the
symmetry energy that involve both symmetric and pure neutron matter. All
these constraints were taken at the saturation point ($\rho_0$), except the
constraints related with the band region. For more details about the constraints and
the criteria used for approval see Ref.~\cite{PRC85-035201}.


The relativistic hadronic models \cite{PRC76-045801,NPA656-331,PRC65-044308}
used here are described by the following Lagrangian densities:

(a) Nonlinear finite range (containing $123$ models):
\begin{small}\begin{eqnarray*}
\mathcal{L}_{\rm NL} = \mathcal{L}_{\rm nm} + \mathcal{L}_\sigma +
\mathcal{L}_\omega
+ \mathcal{L}_\rho + \mathcal{L}_{\sigma\omega\rho},
\label{dl}
\end{eqnarray*}\end{small}
where
\begin{small}\begin{eqnarray*}
\mathcal{L}_{\rm nm} &=& \overline{\psi}(i\gamma^\mu\partial_\mu - M)\psi 
+ g_\sigma\sigma\overline{\psi}\psi 
- g_\omega\overline{\psi}\gamma^\mu\omega_\mu\psi 
- \frac{g_\rho}{2}\overline{\psi}\gamma^\mu\vec{\rho}_\mu\cdot\vec{\tau}\psi, 
\label{lnm} \\
\mathcal{L}_\sigma &=& \frac{1}{2}(\partial^\mu \sigma \partial_\mu \sigma 
- m^2_\sigma\sigma^2) - \frac{A}{3}\sigma^3 - \frac{B}{4}\sigma^4, \\
\mathcal{L}_\omega &=& -\frac{1}{4}F^{\mu\nu}F_{\mu\nu} 
+ \frac{1}{2}m^2_\omega\omega_\mu\omega^\mu 
+ \frac{c}{4}(g_\omega^2\omega_\mu\omega^\mu)^2, \\
\mathcal{L}_\rho &=& -\frac{1}{4}\vec{B}^{\mu\nu}\cdot\vec{B}_{\mu\nu} 
+ \frac{1}{2}m^2_\rho\vec{\rho}_\mu\cdot\vec{\rho}^\mu \qquad\mbox{and} \\
\mathcal{L}_{\sigma\omega\rho} &=& 
g_\sigma g_\omega^2\sigma\omega_\mu\omega^\mu
\left(\alpha_1+\frac{1}{2}{\alpha_1}'g_\sigma\sigma\right)
+ g_\sigma g_\rho^2\sigma\vec{\rho}_\mu\cdot\vec{\rho}^\mu
\left(\alpha_2+\frac{1}{2}{\alpha_2}'g_\sigma\sigma\right) 
\nonumber \\
&+& \frac{1}{2}{\alpha_3}'g_\omega^2 g_\rho^2\omega_\mu\omega^\mu
\vec{\rho}_\mu\cdot\vec{\rho}^\mu.
\label{lomegarho}
\end{eqnarray*}\end{small}

(b) Density dependent models (containing $6$ models):
\begin{small}\begin{eqnarray*}
\mathcal{L}_{\rm DD} &=& \overline{\psi}(i\gamma^\mu\partial_\mu - M)\psi 
+ \Gamma_\sigma(\rho)\sigma\overline{\psi}\psi 
- \Gamma_\omega(\rho)\overline{\psi}\gamma^\mu\omega_\mu\psi 
-\frac{\Gamma_\rho(\rho)}{2}\overline{\psi}\gamma^\mu\vec{\rho}_\mu
\cdot\vec{\tau}\psi \nonumber \\
&+& \frac{1}{2}(\partial^\mu \sigma \partial_\mu \sigma - m^2_\sigma\sigma^2)
- \frac{1}{4}F^{\mu\nu}F_{\mu\nu} + \frac{1}{2}m^2_\omega\omega_\mu\omega^\mu 
-\frac{1}{4}\vec{B}^{\mu\nu}\cdot\vec{B}_{\mu\nu}+\frac{1}{2}m^2_\rho
\vec{\rho}_\mu\cdot\vec{\rho}^\mu,
\end{eqnarray*}\end{small}
where $\Gamma_i(\rho) = \Gamma_i(\rho_0)f_i(x)$, with $f_i(x) =
a_i\displaystyle{\frac{1+b_i(x+d_i)^2}{1+c_i(x+d_i)^2}}$ for $i=\sigma,\omega$,
and $\Gamma_\rho(\rho)~=~\Gamma_\rho(\rho_0)e^{-a(x-1)}$ with
$x=\rho/\rho_0$.

In both Lagrangian densities, $M$, $m_i$, with $i = \sigma, \omega$ and $\rho$
are the nucleon and the mesons masses, respectively. The tensor fields are given by
$F_{\mu\nu} = \partial_\nu\omega_\mu - \partial_\mu\omega_\nu$ and 
$\vec{B}_{\mu\nu}~=~\partial_\nu\vec{\rho}_\mu~-~\partial_\mu\vec{\rho}_\nu$.
$g_i$ ($i = \sigma, \omega$, and $\rho$), $A$, and $B$ are the coupling
constants.

(c) Point-coupling models (containing $18$ models):
\begin{small}\begin{eqnarray*}
\mathcal{L}_{\rm PC} &=& \overline{\psi}(i\gamma^\mu\partial_\mu - M)\psi 
-\frac{\alpha_S}{2}(\overline{\psi}\psi)^2
-\frac{\beta_S}{3}(\overline{\psi}\psi)^3
-\frac{\gamma_S}{4}(\overline{\psi}\psi)^4\\
&-&\frac{\alpha_V}{2}(\overline{\psi}\gamma^\mu\psi)^2
-\frac{\gamma_V}{4}(\overline{\psi}\gamma^\mu\psi)^4 \nonumber 
-\frac{\alpha_{TV}}{2}(\overline{\psi}\gamma^\mu\vec{\tau}\psi)^2
-\frac{\gamma_{TV}}{4}(\overline{\psi}\gamma^\mu\vec{\tau}\psi)^4\\
&-&\frac{\alpha_{TS}}{2}(\overline{\psi}\,\vec{\tau}\psi)^2
+[\eta_1+\eta_2(\overline{\psi}\psi)](\overline{\psi}\psi)(\overline{\psi}
\gamma^\mu\psi)^2 \nonumber 
-\eta_3(\overline{\psi}\psi)(\overline{\psi}\gamma^\mu\vec{\tau}\psi)^2,
\end{eqnarray*}\end{small}
where $\alpha_S$, $\beta_S$, $\gamma_S$, $\alpha_V$, $\gamma_V$,
$\alpha_{TV}$, $\gamma_{TV}$, $\alpha_{TS}$, $\eta_1$, $\eta_2$, $\eta_3$ are
the coupling constants. The subscripts identify the coupling: $S$ stands for
scalar, $V$ for vector, and $T$ for isovector.

The equations of state for these groups, e.g., the energy density and pressure,
are calculated from the energy-momentum tensor (in the mean-field
approximation): $\mathcal{E}~=~\langle T_{00} \rangle$, and $P =
\frac{1}{3}\langle T_{ii} \rangle$ respectively. Other quantities, such as the
symmetry energy, can be obtained from the energy density or the pressure. 

These equations are calculated for each density and proton fraction, $\rho$
and $y = \frac{Z}{A}$ respectively, from the simultaneous solution of the
field equations (obtained from the Euler-Lagrange ones).

The results obtained were the following: out of $147$ models only $9$ of them
satisfies all the constraints, excluding the constraint related to $K_{\tau,\rm
v}$ (isospin dependence of the incompressibility)~\cite{PRC85-035201}. These
are: {\bf {BSR15, BSR16, BSR17}} ($5\%$ of tolerance in the PNM2 constraint --
see ~\cite{PRC85-035201}), {\bf{BSR18}} ($5\%$ of tolerance in the PNM2
constraint -- see~\cite{PRC85-035201})~\cite{PRC76-045801},
{\bf{DD-F,}}~\cite{PRC74-035802} {\bf FSUGold,}~\cite{PRL95-122501} {\bf
FSUGold4,}~\cite{NPA778-10} {\bf FSUGZ06,}~\cite{PRC74-034323} and
{\bf{TW99}}~\cite{NPA656-331}. Their nuclear matter properties are given in
Table~\ref{tab1}. It is important to mention that among these RMF models, $7$
are nonlinear with mixing terms in the $\sigma$ and $\omega$ fields, identified
with a ($*$) and $2$ are density dependent, marked with a ($**$).
\begin{table}[!htb]
\centering
\caption{Properties of nuclear matter at saturation density as
calculated using the RMF models consistent with the macroscopic constraints. All
entries are in MeV.}
\begin{tabular}{lccccccccc}
\hline
{\bf Model} & $\mathbf{\rho_0}$ & $\mathbf{E_0}$ & $\mathbf{K_0}$ &
$\mathbf{m^*}$ & $\mathbf{K^{\prime}}$ & $\mathbf{J}$ & $\mathbf{L}$ &
$\mathbf{K_{\tau,\rm v}}$ \\ \hline
BSR15$^{*}$ & $0.146$ & $-16.03$ & $226.82$ & $0.61$ & $512.29$ & $30.97$ &
$61.79$ & $-252.54$ \\ 
BSR16$^{*}$ & $0.146$ & $-16.05$ & $224.98$ & $0.61$ & $503.17$ & $31.24$ &
$62.33$ & $-258.75$ \\ 
BSR17$^{*}$ & $0.146$ & $-16.05$ & $221.67$ & $0.61$ & $489.45$ & $31.98$ &
$67.44$ & $-287.31$ \\ 
BSR18$^{*}$ & $0.146$ & $-16.05$ & $221.13$ & $0.61$ & $485.73$ & $32.74$ &
$72.65$ & $-318.55$ \\ 
DD-F$^{**}$ & $0.147$  & $-16.04$ & $223.32$ & $0.56$ & $758.73$ & $31.63$ &
$56.00$ & $-285.54$ \\ 
FSUGold$^{*}$ & $0.148$ & $-16.28$ & $229.54$ & $0.61$ & $523.93$ & $32.56$ &
$60.44$ & $-276.07$ \\ 
FSUGold4$^{*}$ & $0.147$ & $-16.40$ & $229.56$ & $0.61$ & $538.33$ & $31.40$ &
$51.74$ & $-205.59$\\ 
FSUGZ06$^{*}$ & $0.146$ & $-16.05$ & $225.06$ & $0.61$ & $503.17$ & $31.18$ &
$62.42$ & $-259.47$ \\ 
TW99$^{**}$ & $0.153$ & $-16.25$ & $240.27$ & $0.55$ & $539.79$ & $32.77$ &
$55.31$ &
$-332.32$ \\ 
\hline
\end{tabular}
\label{tab1}
\end{table}

As in the previous work~\cite{PRC85-035201}, we see that very few models among
the initial 147 satisfy all the constraints used. We note (see Table~\ref{tab1})
that the density-dependent models have values of nucleon effective mass
different from the others. We also observe that if we include in the set
of constraints the ones related to the isospin dependence of the
incompressibility, then none of the models are consistent. Because of this, the
behavior of these models with respect to $K_{\tau,\rm v}$ should be better
investigated.

As a final remark, we highlight the importance of the present study in the
identification of the relativistic parametrizations which better describe the
physics of both nuclear and pure neutron matter.


\begin{theacknowledgments}
M.D. and B.V.C. acknowledge support from FAPESP.
S.S.A., B.V.C., A.D. and D.P.M. acknowledge support
from the CNPq.
%
\end{theacknowledgments}

\bibliographystyle{aipproc}   

\end{document}